\documentclass[pdflatex]{aastex7}

\graphicspath{{./figures/}}

\usepackage{CJK}
\usepackage{amsmath}
\usepackage[utf8]{inputenc}
\usepackage[T1]{fontenc}
\usepackage{anyfontsize}
\usepackage{graphicx}
\usepackage{hyperref}
\submitjournal{ApJ}
\usepackage{gensymb}
\shorttitle{Successive Decimetric Bursts in the Decay Phase of an M-Class Flare}
\shortauthors{Ma et al.}

\begin{document}

\title{Successive Decimetric Bursts Associated with Breakout Magnetic Reconnection during the Decay Phase of an M-Class Flare}

\begin{CJK*}{UTF8}{gbsn}

\author[orcid=0000-0002-5431-6065,gname=Suli,sname=Ma]{Suli Ma}
\affiliation{State Key Laboratory of Solar Activity and Space Weather, National Space Science Center, Chinese Academy of Sciences, Beijing, 100012, China}
\affiliation{National Astronomical Observatories, Chinese Academy of Sciences, Beijing 100101, China}
\email[show]{masuli@nssc.ac.cn}

\author[orcid=0000-0002-7106-6029]{Yihua Yan}
\affiliation{State Key Laboratory of Solar Activity and Space Weather, National Space Science Center, Chinese Academy of Sciences, Beijing, 100012, China}
\affiliation{National Astronomical Observatories, Chinese Academy of Sciences, Beijing 100101, China}
\affiliation{School of Astronomy and Space Science, University of Chinese Academy of Sciences, Beijing, 100049, China}
\email{yanyihua@nssc.ac.cn}

\author[orcid=0000-0003-0747-6450]{Wei Wang}
\affiliation{State Key Laboratory of Solar Activity and Space Weather, National Space Science Center, Chinese Academy of Sciences, Beijing, 100012, China}
\affiliation{National Astronomical Observatories, Chinese Academy of Sciences, Beijing 100101, China}
\affiliation{School of Astronomy and Space Science, University of Chinese Academy of Sciences, Beijing, 100049, China}
\email{wangwei1@nssc.ac.cn}

\author[orcid=0000-0002-6287-3494]{Hamish Reid}
\affiliation{University College London, Mullard Space Science Laboratory, Holmbury Hill Rd, Dorking, 106, Holmbury St Mary, United Kingdom}
\email{hamish.reid@ucl.ac.uk}

\author[orcid=0000-0001-6655-1743]{Ting Li}
\affiliation{State Key Laboratory of Solar Activity and Space Weather, National Space Science Center, Chinese Academy of Sciences, Beijing, 100012, China}
\affiliation{School of Astronomy and Space Science, University of Chinese Academy of Sciences, Beijing, 100049, China}
\email{liting@nssc.ac.cn}

\author[orcid=0000-0002-1682-1714]{Xiaoshuai Zhu}
\affiliation{State Key Laboratory of Solar Activity and Space Weather, National Space Science Center, Chinese Academy of Sciences, Beijing, 100012, China}
\email{zhuxiaoshuai@nssc.ac.cn}

\author[orcid=0000-0001-6076-9370]{Huadong Chen}
\affiliation{National Astronomical Observatories, Chinese Academy of Sciences, Beijing 100101, China}
\affiliation{State Key Laboratory of Solar Activity and Space Weather, National Space Science Center, Chinese Academy of Sciences, Beijing, 100012, China}
\affiliation{School of Astronomy and Space Science, University of Chinese Academy of Sciences, Beijing, 100049, China}
\email{hdchen@nao.cas.cn}

\author{Zhichao Zhou}
\affiliation{State Key Laboratory of Solar Activity and Space Weather, National Space Science Center, Chinese Academy of Sciences, Beijing, 100012, China}
\email{zhouzhichao@nssc.ac.cn}

\author[orcid=0000-0001-9639-2907]{Jing Huang}
\affiliation{National Astronomical Observatories, Chinese Academy of Sciences, Beijing 100101, China}
\affiliation{School of Astronomy and Space Science, University of Chinese Academy of Sciences, Beijing, 100049, China}
\email{huangj@nao.cas.cn}

\author[orcid=0000-0002-3618-3430]{Linjie Chen}
\affiliation{State Key Laboratory of Solar Activity and Space Weather, National Space Science Center, Chinese Academy of Sciences, Beijing, 100012, China}
\affiliation{National Astronomical Observatories, Chinese Academy of Sciences, Beijing 100101, China}
\affiliation{School of Astronomy and Space Science, University of Chinese Academy of Sciences, Beijing, 100049, China}
\email{Chenlinjie@nssc.ac.cn}

\author[orcid=0000-0001-5359-6897]{Chengming Tan}
\affiliation{State Key Laboratory of Solar Activity and Space Weather, National Space Science Center, Chinese Academy of Sciences, Beijing, 100012, China}
\affiliation{National Astronomical Observatories, Chinese Academy of Sciences, Beijing 100101, China}
\affiliation{School of Astronomy and Space Science, University of Chinese Academy of Sciences, Beijing, 100049, China}
\email{tanchengming@nssc.ac.cn}

\author[orcid=0000-0002-1810-6706]{Xingyao Chen}
\affiliation{State Key Laboratory of Solar Activity and Space Weather, National Space Science Center, Chinese Academy of Sciences, Beijing, 100012, China}
\email{chenxingyao@nssc.ac.cn}

\author[orcid=0000-0003-1912-2959]{Minghui Zhang}
\affiliation{State Key Laboratory of Solar Activity and Space Weather, National Space Science Center, Chinese Academy of Sciences, Beijing, 100012, China}
\email{zhangminghui@nssc.ac.cn}

\author[orcid=0000-0001-8988-2350]{Xin Yao}
\affiliation{State Key Laboratory of Solar Activity and Space Weather, National Space Science Center, Chinese Academy of Sciences, Beijing, 100012, China}
\email{yaoxin@nssc.ac.cn}

\begin{abstract}
Radio imaging spectroscopy diagnoses coronal electron acceleration and propagation by tracking energetic electron beams and locating energy-release sites. Using MUSER (0.4-2 GHz) and SDO data, we investigate decimetric radio emission during an eruption with an M1.1 flare on 17 December 2014. The eruption involves a twisted coronal structure evolving through three stages: a violent internal eruption, a rise-and-expansion phase, and a breakout eruption. The flare peaks at 01:50 UT during the internal eruption, while the cluster of successive type III-like decimetric bursts occurs in the decay phase, linked to the breakout eruption. The radio source coincides with a high-squashing-factor Q region from nonlinear force-free field (NLFFF) extrapolations, indicating a quasi-separatrix layer (QSL). We interpret the radio emission as consistent with plasma emission driven by nonthermal electrons accelerated by breakout reconnection at this QSL, between the outer envelope of the erupting structure and the surrounding corona. The successive bursts occur in the decay phase because breakout reconnection involves a larger, more energetic twisted structure at higher altitude, where an extended high-Q region, favorable acceleration geometry, and repeated sub-eruptions sustain efficient electron acceleration. Missing soft X-ray enhancement suggests most accelerated electrons escape upward or lose energy before reaching the chromosphere. The concentrated source positions of the successive radio bursts suggest pseudostreamers may act as particle accelerators, trapping electrons and directing them along a narrow layer. These results establish decimetric radio imaging as a powerful diagnostic of breakout reconnection and particle acceleration in the solar corona, with implications for energy release in pseudostreamer environments.
\end{abstract}

\keywords{Sun: activity; Sun: corona; Sun: flares; Sun: radio radiation; Sun: magnetic fields; techniques: interferometric; radiation mechanisms: non-thermal; acceleration of particles}

\section{Introduction}
Solar flares and coronal mass ejections (CMEs) are driven by the impulsive release of magnetic free energy stored in the solar corona. This rapid energy release accelerates particles to high energies and produces enhanced emission across the electromagnetic spectrum \citep{Benz+2017}; when directed toward Earth, such events can cause significant space weather disturbances.
It is generally accepted that magnetic reconnection (MR) is a ubiquitous process that transforms magnetic energy into both heat and kinetic energy \citep[e.g.,][]{Priest+2000,Forbes+2006,Chen+2011,Shibata+2011,Fletcher+2011,Schmieder+2015,Jiang+2024,liting+2021,Dudik+2025}.
Although early theoretical models were largely two-dimensional, the coronal magnetic field is intrinsically three-dimensional. Three-dimensional (3D) reconnection exhibits fundamentally different properties: it can occur without null points, involves a finite diffusion region, and is characterized by the continuous slipping or flipping of field lines \citep{Pontin+2022}. Direct observational detection of the thin current sheets where energy release takes place remains challenging \citep{Linj+2015}. Quasi-separatrix layers (QSLs) have become a standard framework for interpreting 3D reconnection, with observations of slipping footpoints and ribbon motions verifying theoretical predictions \citep{Gou+2016, Dudik+2025}.
Besides the QSLs, 3D null-point is another important location for 3D reconnection, especially during the circular ribbon (CR) flares \citep{Masson+2009}. The 3D null-point reconnection in the fan-spine topology is similar to the “breakout” nature \citep[e.g.,][]{Antiochos+1999,Sunxd+2013}.  
High-resolution multiwavelength observations of such events reveal sequential brightening and slipping motions along the ribbons, consistent with 3D reconnection in the fan-spine topology \citep{Wanghm+2012, Liting+2018, Huang+2024, Yang+2024}.

A critical consequence of 3D reconnection is the efficient acceleration of nonthermal electrons. As the primary site of energy release, the reconnection current sheet converts magnetic energy into the kinetic energy of energetic electrons\citep[e.g.,][]{Shen+2011,Mei+2012,LiX+2025,Yan+2022}. These accelerated electrons subsequently propagate along field lines and generate radio emission, offering a direct diagnostic of both the acceleration and transport processes in the solar corona\citep{Duan+2022, Karlicky+2025}. Spanning from gamma-ray to kilometer wavelengths, such emission provides a unique suite of diagnostic tools for solar eruptions \citep[e.g.,][]{Wild+1963, Dulk+1985, Hudson+1995, Bastian+1998, Pick+2008, Gary+2023}.
At millimeter and centimeter wavelengths, incoherent gyrosynchrotron emission from relativistic electrons dominates the radio output, exhibiting a close correspondence with X-ray emission during the impulsive phases of solar flares. In contrast, at decimeter and meter wavelengths, coherent plasma radiation from near-relativistic electrons becomes prevalent \citep{Melrose+1985}. \citet{Benz+2005} presented a statistical analysis and found that decimetric type III bursts, as well as pulsations, spikes, and continua, tend to be detected preferentially during the peak hard X-ray phases of solar flares, but not consistently enough to imply a common origin. They further suggested that the lack of detailed correlation between hard X-ray and radio emissions points to the likelihood of multiple, possibly independent, acceleration mechanisms operating during flares. Recently, \citet{Paipaleon+2025} surveyed the relationship between hard X-ray and radio emissions and supported multiple reconnection episodes.

In addition to emissions observed at the Sun, electrons impulsively accelerated to energies of $\sim$1\sbond100 keV are routinely detected in interplanetary space \citep[e.g.,][]{Lin+1985}. Furthermore, \citet{Lin+1985} found that approximately 90\% of low-energy impulsive electron events coincide with interplanetary type III radio emission in the 30~kHz to a few MHz range; about 50\% of these also correlate with metric or decimetric type III bursts detected at frequencies above $\sim$20~MHz by ground-based observatories. \citet{Wang_lh+2012} surveyed 1191 solar electron events and found that they have a 98.75\% association with type III radio bursts. \citet{Benz+2001} statistically traced the origins of energetic electrons and found that most events (22 out of 27, or 81\%) are associated with metric (coronal) III bursts, and the majority of these bursts have starting frequencies between 150 and 450~MHz.

Due to the low density at and above the height corresponding to these radio emissions, studying the acceleration of electrons or MR with observations at other wavelengths, such as EUV and X-ray, is challenging. Therefore, radio emission at metric or decimetric wavelengths plays an important role in understanding the acceleration of electrons and their propagation in the middle or high corona.

On 17 December 2014, the Mingantu Spectral Radioheliograph \citep[MUSER;][]{Yanyh+2009, Yanyh+2016, Yanyh+2021, Yanyh+2023} captured a series of type III-like bursts in the decay phase of the SOL2014-12-17T01:50 M1.1 flare at 400~MHz--2~GHz. Based on data from MUSER and the Atmospheric Imaging Assembly \citep[AIA;][]{Lemen+2012} aboard the Solar Dynamics Observatory \citep[SDO;][]{Pesnell+2012}, we performed a detailed analysis of the solar eruption. We present a brief introduction to the data in Section~2. Observations and results are presented in Section~3, and the discussion and conclusion are in Section~4.

\section{Data}
MUSER is a new generation of solar-dedicated radio imaging-spectroscopic telescope covering the wideband range of 30~MHz--15~GHz \citep{Yanyh+2025}. To date, it has been extended to three sets of antenna arrays covering frequency ranges of 30--400~MHz \citep[MUSER-L]{Yanyh+2025}, 0.4--2~GHz (MUSER-I), and 2--15~GHz (MUSER-H, also referred to as MUSER-II). Each array is distributed along three spiral arms, with the longest baseline measuring 3.0~km. This study utilizes both spectroscopic and imaging data from the MUSER-I instrument. MUSER-I comprises 40 antennas, each with a diameter of 4.5~m, operating across a frequency range of 0.4--2~GHz. It provides data with a temporal resolution of 25~ms and a frequency resolution of 25~MHz. 

Data processing for MUSER is based on standard aperture synthesis imaging techniques \citep[e.g.,][]{Taylor+1999,Thompson+2001}. General pipelines for MUSER data processing were established by \citet{wangw+2013}. Subsequent work developed and refined position calibration methods utilizing satellites at 1.7~GHz or intense solar bursts \citep{wangw+2013, wangw+2019, Wangw+Yanyh_2019}. Alternative methods employing the solar disk for calibration have also been explored \citep{Chenlj+2017, Chenxy+2019, Mei+2018, wangw+2019}, with a recent advancement presented for enhanced spatial calibration by \citet{Zhouzc+2022}. The intensity calibration methodology for MUSER (covering 0.4--15~GHz) is detailed in \citet{Zhouzc+2022}.

Statistical studies of radio burst events observed spectroscopically by MUSER-I during its 2014--2019 operational period have been presented by \citet{zhangmh+2021}. Additional case studies employing either spectroscopic or imaging data from MUSER-I include works by \citet{Chenxy+2019, Hong+2021, Huangj+2022}.

In this study, the final calibrated MUSER-I images were produced through the following procedure: (1)~for each frequency channel, an intense radio burst was selected as a calibrator to facilitate initial phase calibration; (2)~dirty images were generated by applying a Fourier transform to the visibility data using this calibrator; (3)~additional phase calibration was applied following the method described in \citet{Zhouzc+2022}; and (4)~the CLEAN algorithm \citep{Hogbom_1974, wangw+2019} was performed to produce the final clean images.

The positional uncertainty at each frequency is estimated based on the spatial resolution corresponding to the longest baseline utilized. For the observations presented here, the maximum baseline is approximately 1.3~km. This yields an approximate resolution of $\sim$130\arcsec\ at 0.4~GHz and $\sim$50\arcsec\ at 1.4~GHz along the minor axis of the synthesized beam. Consequently, the absolute positional accuracy is constrained by these resolution limits. However, the relative positional alignment between images acquired at different times but at the same frequency is significantly more precise, being governed primarily by the pointing stability of the instrument.

Additionally, a dynamic spectrum was obtained from the cross-correlation data using the same baselines employed for imaging. This method provides greater intensity sensitivity compared to single-antenna spectra.

AIA provides 1.5-arcsec spatial and 12-second temporal resolution coronal images in seven EUV bandpasses, and we mainly used 94~\AA\ (6.8), 131~\AA\ (5.6, 7.0), 171~\AA\ (5.8), and 211~\AA\ (6.3). Parentheses list the characteristic temperature in logarithmic scale, $\log T$, for each line. Magnetograms from the Helioseismic and Magnetic Imager \citep[HMI;][]{Scherrer+2012} are used to display the magnetic field topology. The Nobeyama Radioheliograph \citep[NoRH;][]{Nakajima+1994} at 17~GHz is used to show microwave emission, which can be used to trace streaming of energetic electrons from particle acceleration sites. The Ramaty High Energy Solar Spectroscopic Imager \citep[RHESSI;][]{Lin+2002} is designed to investigate particle acceleration and energy release in solar flares. Only low-energy passbands ($<$25~keV) show signals during this eruption.

\section{Observations and Results}
A series of decimetric bursts was captured by MUSER-I from 01:40 to 02:30~UT on 17 December 2014, with the most violent emission occurring at approximately 02:12~UT, in the deep decay phase of its associated M1.1 flare. Utilizing radio dynamic spectra and images from MUSER-I, EUV images from AIA, microwave data at 17~GHz from NoRH, and hard X-ray data from RHESSI, we present a comprehensive analysis of successive decimetric/metric solar radio bursts during 01:40--02:30~UT and their associated eruption of a complex twisted structure.

\subsection{Overview of the Event}

\begin{figure}[ht!]
\includegraphics[width=1\textwidth]{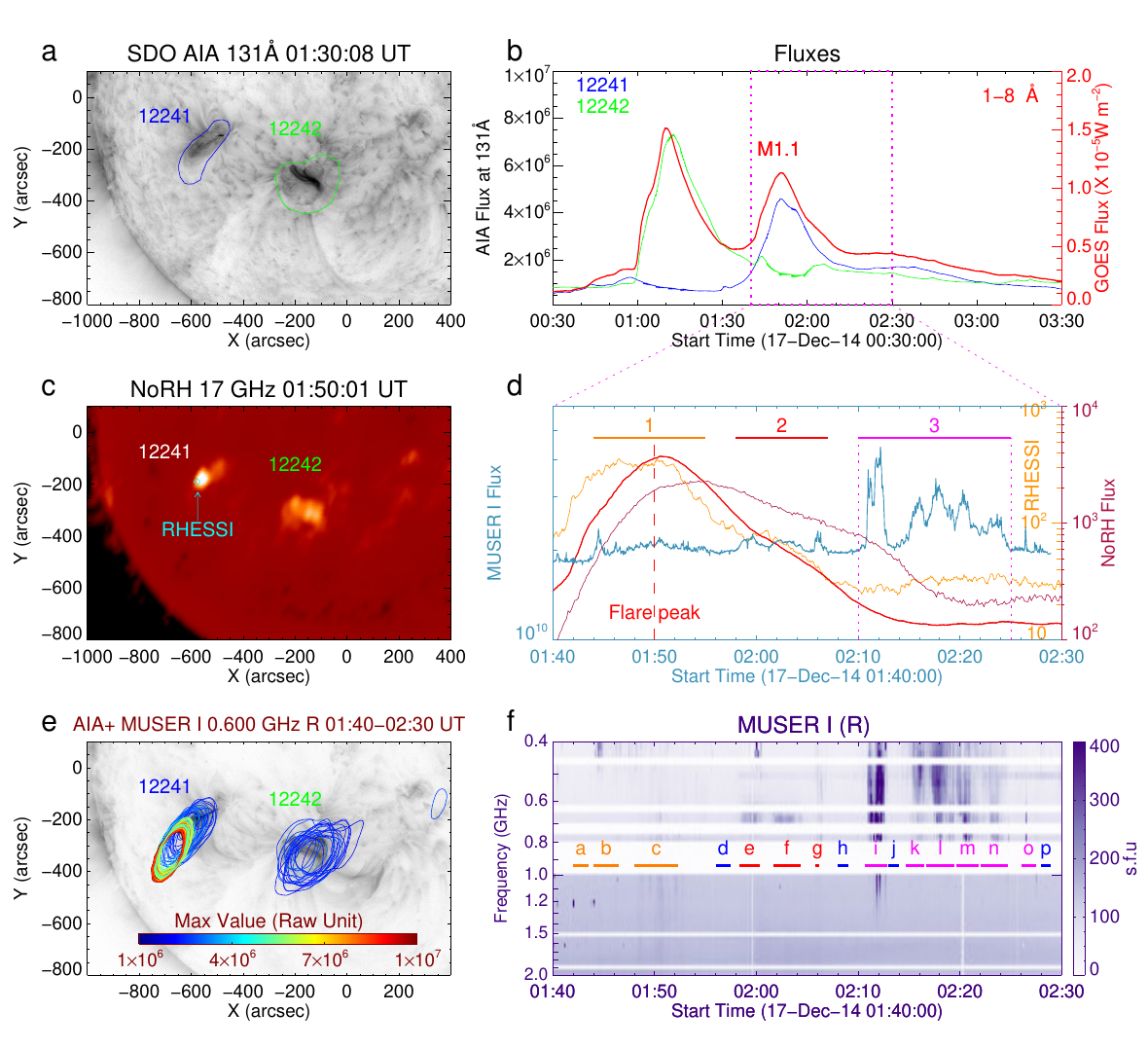}
\caption{The source region of the M1.1-class flare and its surrounding environment. a. The SDO/AIA 131~\AA\ image presents EUV observations of AR 12241 and AR 12242, with blue and green closed curves delineating the areas used for flux measurements. b. The time profiles show the GOES total solar disk flux in red and the 131~\AA\ flux for AR 12241 in blue and AR 12242 in green. c. NoRH 17~GHz image with RHESSI contours at 6--12~keV. d. Time profiles of AR 12241 from MUSER-I in turquoise, NoRH in dark purple, and RHESSI 3--25~keV in orange. The red profile is the GOES 1--8~\AA\ full-disk flux profile. The bars labeled `1', `2', and `3' indicate different eruption stages. e. Radio source contours at 80\% of the maximum of MUSER images from 01:40 to 02:30~UT in right circular polarization, overlaid on the SDO/AIA 94~\AA\ image. f. MUSER-I cross-correlated dynamic spectrum displays decimetric radio bursts associated with the M1.1-class flare in right circular polarization. Short bars labeled with letters indicate the integration intervals for the MUSER-I burst sources, as illustrated in Figures~\ref{fig:contour} and~\ref{fig:centroid}.}
\label{fig:fig1}
\end{figure}

Two active regions, AR 12241 (S11E37) and AR 12242 (S19W02), are highly active during this period. Since the GOES X-ray flux represents the integrated signal from the entire disk and cannot resolve individual region contributions, we instead utilize the AIA 131~\AA\ flux as a proxy, as both diagnostics trace plasma at temperatures above 10~MK \citep[e.g.,][]{Woods+2011, Chenbin+2014}, to determine their contributions. Figure~\ref{fig:fig1}a displays the 131~\AA\ image, with blue and green curves outlining the regions used to extract fluxes from AR 12241 and AR 12242, respectively. The corresponding fluxes, shown in Figure~\ref{fig:fig1}b with matching color coding, are plotted alongside the GOES flux (in red). This comparison demonstrates that the M1.1 flare originates predominantly in AR 12241, with a smaller contribution from AR 12242. The flux and dynamic spectrum from 01:40 to 02:30~UT are presented in Figures~\ref{fig:fig1}d and~\ref{fig:fig1}f, covering the interval marked by the purple dashed lines in Figure~\ref{fig:fig1}b, based on MUSER observations starting at 01:40~UT.

Figure~\ref{fig:fig1}c shows the NoRH image at 17~GHz near the flare peak, revealing comparatively stronger emission from AR 12241 and weaker emission from AR 12242. Figure~\ref{fig:fig1}e overlays MUSER-I contours at 0.6~GHz on the AIA 94~\AA\ image; contours are drawn at 80\% of the maximum intensity, and the scale reflects the peak values. The results show that stronger energy release mainly comes from AR 12241, while weaker emission comes from both regions.

We calculate the fluxes from AR 12241 at multiple wavelengths/frequencies, as presented in Figure~\ref{fig:fig1}d. The turquoise, dark purple, and orange curves correspond to fluxes measured by MUSER-I (0.675~GHz), NoRH (17~GHz), and RHESSI (6--12~keV), respectively, while the full-disk GOES 1--8~\AA\ flux is included for comparison. Notably, intense decimetric radio bursts are observed during the decay phase of the flare, exhibiting a temporal behavior distinct from the RHESSI X-ray and NoRH fluxes, which closely follow the trend of the GOES flux. Figure~\ref{fig:fig1}f shows the dynamic spectrum derived from the total cross-power in right circular polarization recorded by MUSER-I. Based on the temporal evolution and spectral properties of the decimetric radio emission, we classified the eruption into three stages. Stage~1 exhibits low flux density with a broad frequency passband and enhanced pulsations. Stage~2 maintains low flux density but is confined to a narrow frequency band. Stage~3 is distinguished by elevated flux density and pronounced high-frequency cutoffs. The time intervals corresponding to these three stages are indicated by horizontal lines labeled 1--3 in Figure~\ref{fig:fig1}d. Short bars labeled with letters in Figure~\ref{fig:fig1}f indicate the time intervals of the MUSER-I burst sources discussed in the following text.

\subsection{Eruption of the Complex Twisted Structure}

\subsubsection{Morphology of the Complex Twisted Structure}

\begin{figure}[ht!]
\includegraphics[width=1\textwidth]{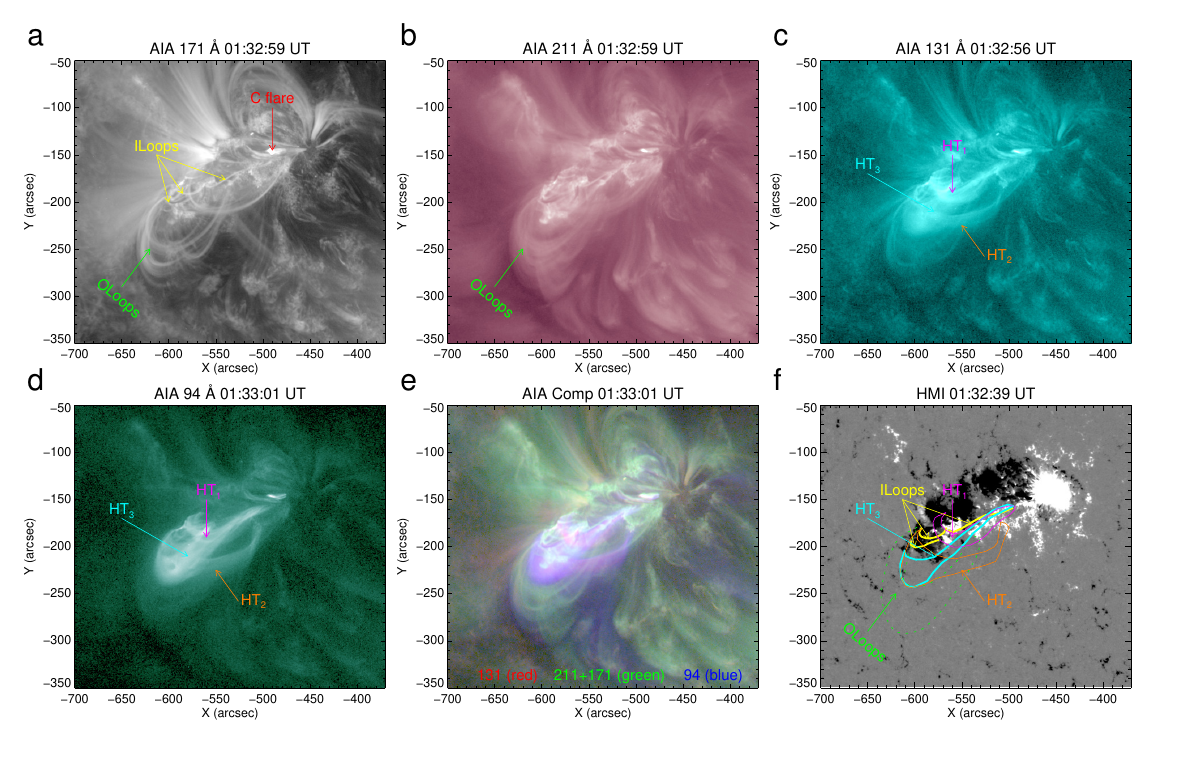}
\caption{Multi-wavelength EUV images and the corresponding photospheric magnetogram reveal the morphology of the twisted coronal structure. The ``ILoops'' and ``OLoops'' denote the inner and outer loops, while ``HT$_1$'', ``HT$_2$'', and ``HT$_3$'' represent different parts of the hot twisted structure. The red arrow marks a C-class flare that likely initiated the eruption, resulting in the M-class flare.}
\label{fig:morphology}
\end{figure}

Figure~\ref{fig:morphology} presents SDO/AIA EUV observations of the complex twisted structure at multiple wavelengths and the corresponding SDO/HMI photospheric magnetogram at 01:33~UT, at the onset of the eruption that led to the M1.1 flare. Figures~\ref{fig:morphology}a--d display AIA images of the erupting region in 171~\AA, 211~\AA, 131~\AA, and 94~\AA, respectively, with dominant temperatures of approximately 0.6, 2, 0.4/12, and 6.3~MK. In Figure~\ref{fig:morphology}a, the ``ILoops'' denote the inner cooler loops that become part of the twisted structures involved in the subsequent eruption, while the ``OLoops'' denote the overlying loops predominantly visible in the 171~\AA\ and 211~\AA\ passbands. The ``HT$_1$'', ``HT$_2$'', and ``HT$_3$'' features in the 131~\AA\ and 94~\AA\ channels represent distinct parts of the hot twisted structure. Figure~\ref{fig:morphology}e shows an RGB composite of these channels (131~\AA\ in red, 211~\AA\ + 171~\AA\ in green, and 94~\AA\ in blue), depicting the spatial distribution of the cool and hot structures, which together constitute the entire twisted erupting structure. The red arrow in Figure~\ref{fig:morphology}a marks a C-class flare that likely triggers the loss of equilibrium of the twisted structure, leading to subsequent eruptions. Figure~\ref{fig:morphology}f shows the corresponding HMI line-of-sight magnetogram overlaid with contours outlining the ``ILoops'', ``OLoops'', ``HT$_1$'', ``HT$_2$'', and ``HT$_3$'', revealing the magnetic footpoint locations of these structures.

\subsubsection{Magnetic Topology and Logarithmic \texorpdfstring{$Q$}{Q}}

\begin{figure}[ht!]
\includegraphics[width=1\textwidth]{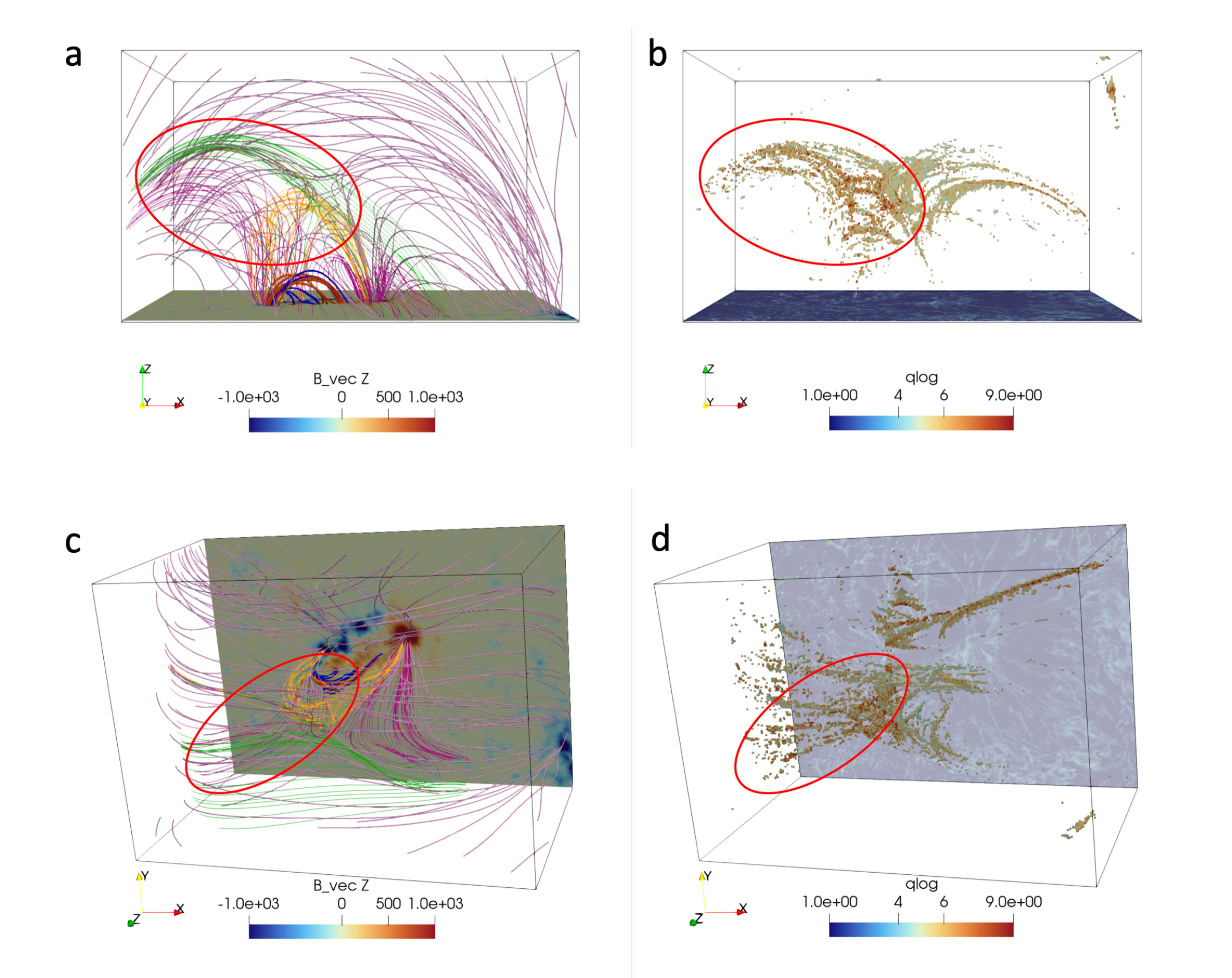}
\caption{Spatial distribution of the magnetic topology and logarithmic squashing factor \(Q\) around AR 12241 from NLFFF modeling. Panels a and b show side views of the extrapolated field lines and \(\log Q\), respectively; panels c and d show the corresponding slightly top-down views. Thin lines of various colors in panels a and c represent large-scale closed loops or open field lines, while thick lines highlight the inner twisted structure central to the eruption. Red ellipses mark the high-\(Q\) regions associated with the successive radio bursts.}
\label{fig:extrapolation}
\end{figure}

To investigate the magnetic topology underlying the eruption, we perform a nonlinear force-free field (NLFFF) extrapolation of AR 12241 using the photospheric vector magnetogram obtained by SDO/HMI, following the method of \citet{Wiegelmann2004}. The extrapolated field lines reveal a complex magnetic configuration composed of distinct structural systems, which we distinguish by color and line thickness. Thin lines in various colors represent large-scale closed loops or open field lines, while the thick lines highlight the inner twisted structure central to the eruption. Although the NLFFF extrapolation captures the overall topology of this twisted structure, reconstructing a true flux rope remains challenging. As shown by \citet{Leake+2026}, the photospheric magnetogram alone may not fully recover its complex three-dimensional structure, especially in non-equilibrium states. We therefore regard the extrapolated field lines as a topological guide only. This field is then used to compute the squashing factor \(Q\) following the methods of \citet{chen+2026} \citep[see also][]{Titov+2002,Liu+2016}, which quantifies the distortion of the field-line mapping and serves as a robust diagnostic for quasi-separatrix layers and separatrix surfaces \citep{Priest+1995, Demoulin+1996, Wilmot-Smith+2009}. Figures~\ref{fig:extrapolation}a and~\ref{fig:extrapolation}c show side and slightly top-down views of the extrapolated field lines, respectively, revealing a complex, twisted magnetic configuration. Figures~\ref{fig:extrapolation}b and~\ref{fig:extrapolation}d display the corresponding logarithmic \(Q\) distributions for the same two viewing angles. The red ellipsoids highlight only those high-\(Q\) regions indicative of strong field-line divergence, characteristic of quai-separatrix layer and preferred sites for magnetic reconnection \citep[e.g.,][]{Titov+2002, Aulanier+2005, Pontin+2024, Kang+2017, Shen+2019}, which generally correspond to the locations of the successive radio bursts examined in detail in the following sections.

\begin{figure}[ht!]
\includegraphics[width=1\textwidth]{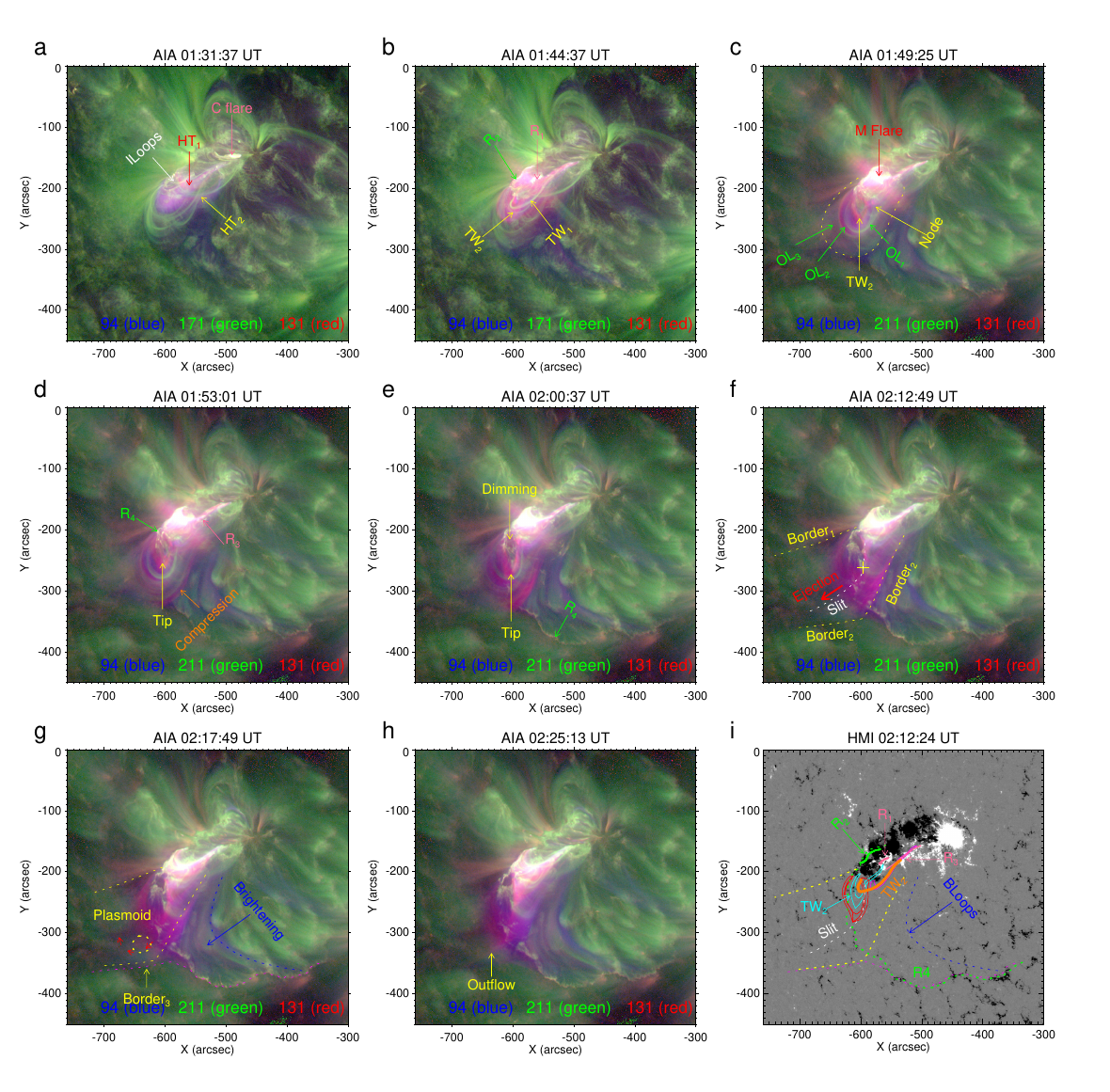}
\caption{EUV observations illustrate the eruption process of the complex twisted structure. Panels a--h present composite-color images combining 131~\AA\ (red), 171~\AA\ (green), and 94~\AA\ (blue) wavelengths. Panel~i shows the photospheric magnetic field observed by HMI. The pink arrow in panel a marks the ``C flare'', which likely acts as the trigger, while ``HT$_1$'' and ``HT$_2$'' denote two hot parts of the twisted structures, and the white arrow labeled ``ILoops'' identifies the inner short sheared loops. ``$R_1$'', ``$R_2$'', ``$R_3$'', and ``$R_4$'' in panels b, d, e, and i refer to different flare ribbons. ``TW$_1$'' and ``TW$_2$'' in panels b, c, and i indicate the inner cold and hot twisted structures, respectively. ``OL$_1$'', ``OL$_2$'', and ``OL$_3$'' label the overlying coronal loops. ``M flare'' indicates the main location of the M1.1 flare. The ``Node'' marks the position where parts of the twisted structure twist together to form a node. The ``Tip'' in panels d and e marks the tip of ``TW$_2$'', which is likely stretched by ``TW$_1$''. The ``Compression'' in panel d indicates the area showing distinct compression between the erupting structure and surrounding magnetic loops. The ``Dimming'' in panel e shows a deep dimming area at the negative footpoint, which indicates the opening of the coronal magnetic structure and material escaping. ``Ejection'' with a red arrow in panel f indicates the general projection direction of the eruption, with ``Border$_1$'' and ``Border$_2$'' marking its borders. The ``slit'' with a white dashed curve indicates the slit along which we obtained the time-slit image. The ``Plasmoid'' in panel g indicates a rotated plasmoid that appeared during the eruption, with three arrows indicating its rotation direction. The ``Brightening'' points to the brightening area, while ``Border$_3$'' with dashed pink curves marks its southern border, which corresponds to the semicircular flare ribbon. The ``Outflow'' in panel h marks the outflow at the end of the eruption. An animation of the eruption from 01:00 to 03:00~UT is available. It displays the temporal evolution in single-wavelength and composite-color images, illustrating the dynamic behavior of the structures shown in the static panels. }
\label{fig:fig2}
\end{figure}

\subsubsection{Disappearance of Inner Loops and the M1.1 Flare}

Figures~\ref{fig:fig2}a--h display the eruption process of the twisted structure. To display the multi-temperature eruption structure simultaneously, composite-color EUV images are used: red and blue indicate the hot 131~\AA\ and 94~\AA\ channels, respectively, and green represents the cool 171~\AA\ channel (Figures~\ref{fig:fig2}a--b) or the warm 211~\AA\ channel (Figures~\ref{fig:fig2}c--h). The photospheric magnetic field is displayed in Figure~\ref{fig:fig2}i to aid in understanding the magnetic connectivity of the erupting coronal structure, with white for positive polarity and black for negative polarity.

Figure~\ref{fig:fig2}a illustrates the onset of the eruption of the complex twisted structure. The hot parts of this structure, such as ``HT$_1$'' and ``HT$_2$'', exhibit upward motion following the appearance of a compact C-class flare in the northern portion of AR 12241, which marks the start of the rising phase in the 131~\AA\ flux (see Figure~\ref{fig:fig1}b). Meanwhile, some of the low coronal loops, indicated as ``ILoops'' and mainly appearing in cool lines such as 171~\AA, begin to disappear, and simultaneously a flare with two short ribbons, marked ``$R_1$'' and ``$R_2$'' in Figure~\ref{fig:fig2}b, appears.

As the eruption evolves, a thin twisted coronal structure (TW$_1$), possibly a flux rope and distinct in the cool 171~\AA\ passband, and a thicker, hotter twisted coronal structure (TW$_2$), visible in the hot 131~\AA\ and 94~\AA\ channels, appear and move outward, driving the overlying coronal structures outward and compressing the surrounding corona. Figure~\ref{fig:fig2}b shows the relative morphologies and positions of TW$_1$ and TW$_2$: the southeast section of TW$_1$ is anchored in a negative magnetic field beneath TW$_2$, while its northwest segment lies above TW$_2$. The rapid upward movement of TW$_1$ pushes TW$_2$ outward. By about 01:47~UT, TW$_1$ fades from view in the cooler passband, whereas the M1.1 flare peaks at 01:50~UT. Although the two are likely temporally related, no clear spatial link between them is evident.

Meanwhile, TW$_2$ is stretched and forms a cusp at its top (see Figures~\ref{fig:fig2}c--e). A dotted bright flare ribbon labeled ``$R_3$'' in Figure~\ref{fig:fig2}d appears near the positive twisted feet of TW$_1$ and TW$_2$. Meanwhile, ``$R_4$'' in Figure~\ref{fig:fig2}d appears in the negative feet of these structures and spreads outward to form a semicircle. Notably, the positive section of TW$_2$ becomes entangled with other loop structures, creating a node labeled ``Node'' in Figure~\ref{fig:fig2}c, which superficially resembles a false footpoint. As TW$_2$ migrates southward, the node appears to stay anchored in place, indicating that multiple twisted or tangled loops are intertwined.

With the ascent of TW$_2$, several layers of alternating cool and hot coronal loops become visible from inner to outer, labeled ``OL$_1$'', ``OL$_2$'', and ``OL$_3$'' in Figure~\ref{fig:fig2}c. As the outer part of TW$_2$ ascends, the inner overlying warm loops ``OL$_1$'' progressively contract and eventually disappear. Meanwhile, the outer coronal loops ``OL$_3$'' expand southward, pressing against the boundary of an adjacent loop system---the separatrix layer \citep{Priest+2003}---and producing a high-temperature region, as indicated by the orange arrow in Figure~\ref{fig:fig2}d. Accompanying the expansion of the outer loops, a semicircular flare ribbon appears and becomes pronounced after 01:56~UT. This feature is highlighted with a green arrow and labeled ``$R_4$'' in Figure~\ref{fig:fig2}e, which is an elongated part of the original $R_4$. Along with this, a brightening enhancement is identified in 94~\AA, as indicated in Figure~\ref{fig:fig2}g. The brightening area, primarily observed at 94~\AA, can be traced back to the early expansion of the tangled structure (see the animation attached with Figure~\ref{fig:fig2}).

At approximately 02:12~UT, a violent eruptive phase began, initiated by a break in the positive section of TW$_2$, indicated by a yellow plus symbol in Figure~\ref{fig:fig2}f. This was followed by the outward expulsion of overlying coronal loops and associated material, predominantly along the path delineated by the dashed yellow line with a red arrow, which also served as the reference for a time-slit image. During this expulsion, the magnetic field displayed dynamics similar to those of a cracking whip. Several minutes later, a plasmoid appeared, highlighted by a small dashed yellow circle in Figure~\ref{fig:fig2}f, with its subsequent motion traced by red arrows. The dashed yellow curves in Figures~\ref{fig:fig2}f--g indicate the borders of the eruption and are labeled ``Border$_1$'' and ``Border$_2$'' in Figure~\ref{fig:fig2}f. The magenta dashed curve marked ``Border$_3$'' indicates the northern border of the enhanced brightening area indicated with the blue ``Brightening'' and the arrow in Figure~\ref{fig:fig2}g. Close to the end of the eruption, a narrow outflow along the spine was observed in the hot EUV channel (Figure~\ref{fig:fig2}h).

Figure~\ref{fig:fig2}i shows the HMI magnetogram, overlaid with the flare ribbons ``$R_1$'' and ``$R_2$'' in the early eruption stage (Period~1). TW$_1$, TW$_2$, ``$R_3$'', and ``$R_4$'', as well as the borders, indicate a possible separatrix or quasi-separatrix.

\subsection{Eruption in Radio Observation}

\begin{figure}[ht!]
\includegraphics[width=0.7\textwidth]{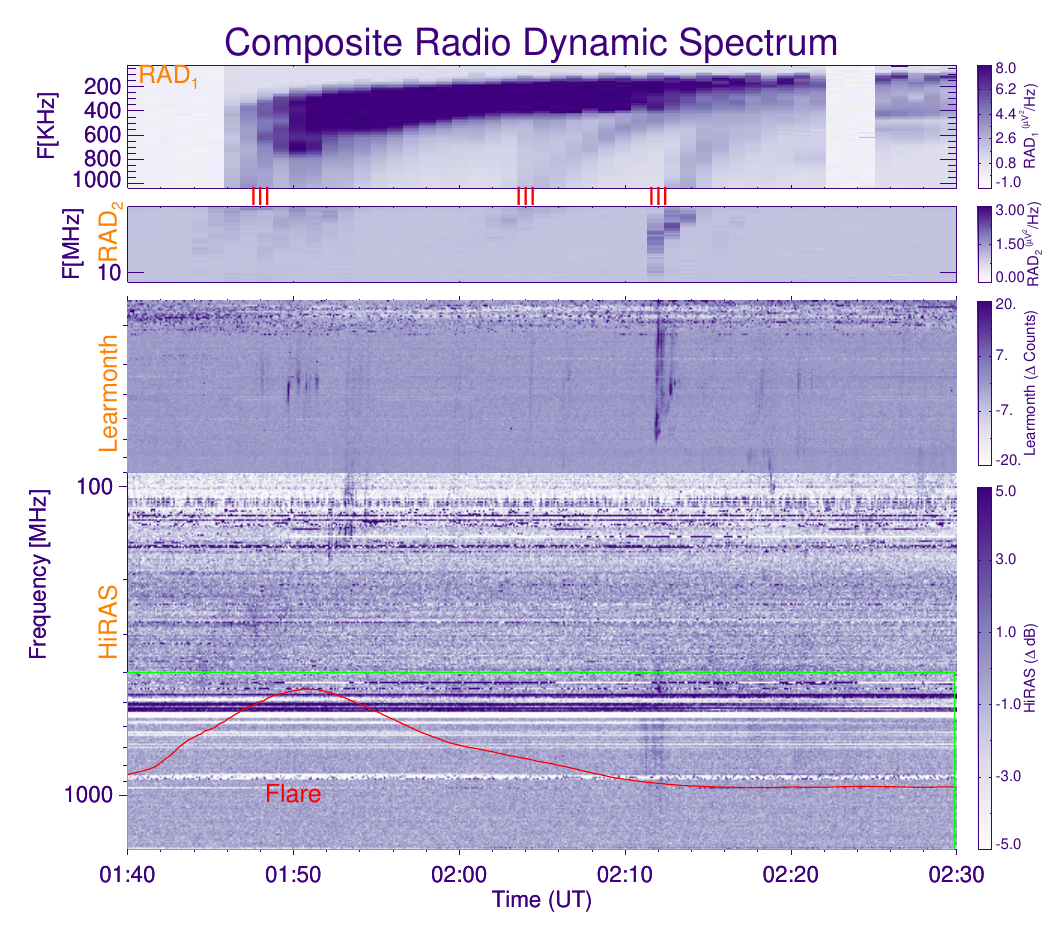}
\caption{The composite spectrum, incorporating data from HiRAS, Learmonth, WIND/RAD$_1$, and WIND/RAD$_2$, reveals a broadband radio eruption extending from \(\sim\)2~GHz down to a few kHz. The color bars on the right indicate the intensity scales: power spectral density in \(\mu\)V\(^2\)/Hz for the WIND data, and background-subtracted intensities in \(\Delta\)dB and \(\Delta\)Counts for HiRAS and Learmonth, respectively. The red curve denotes the corresponding 1--8~\AA\ flux measured by GOES. The green box delineates the frequency and time range observed by MUSER-I, which is shown in detail in Figure~\ref{fig:fig1}f.}
\label{fig:spectrum}
\end{figure}

\subsubsection{Broadband Radio Spectrum of the Eruption}

To contextualize this fine-scale MUSER-I activity within the full broadband event, we present a composite radio spectrum in Figure~\ref{fig:spectrum}. The complete eruption commenced at 01:40~UT with weak, sporadic bursts---both isolated and clustered---in the decimeter and meter wavebands. This early activity was followed by several interplanetary type III bursts observed by WIND/RAD\(_1\) and RAD\(_2\), suggesting the extension of the metric emission into the kilometric domain. The spectrum displays the background-subtracted intensity in \(\Delta\)dB and \(\Delta\)Counts for the ground-based instruments, and the power spectral density in \(\mu\)V\(^2\)/Hz for WIND. The timeline of the event is referenced against the 1--8~\AA\ GOES soft X-ray flux (red curve). The specific observing window of MUSER-I, corresponding to the dynamic spectrum in Figure~\ref{fig:fig1}f, is demarcated by the green box.

This composite view reveals how the three evolutionary stages identified in MUSER-I data relate to the broader emission. During Stage~1, the dominant radio power resides below 500~MHz, and MUSER-I captured only the high-frequency segment of drifting pulsations shifting from 500~MHz to 100~MHz between 01:44 and 01:56~UT , which is highlighted in Figure~\ref{fig:spectrum}. The broadband radio burst around the flare peak, which can be identified in the MUSER cross-power spectrum in Figure~\ref{fig:fig1}f, is not clear in the composite image, indicating that the cross-power spectrum is more sensitive than a single antenna to weaker signals. The narrowband emission defining Stage~2 is centered at 600--800~MHz, as clearly seen in Figures~\ref{fig:sp_polar}c and~\ref{fig:sp_polar}d. Several type III bursts are also identified in the 20--200~MHz range during Stage~2, visible in Figure~\ref{fig:spectrum}, with some extending down to 400~kHz. Finally, Stage~3, characterized by elevated flux density, appears as a bifurcation of the emission into two components: a low-frequency component below 100~MHz, discernible in Figure~\ref{fig:spectrum} and a distinct, high-intensity component above 300~MHz, the latter being the strongest emission detected by MUSER-I.

\subsubsection{Representative Spectral Features from MUSER-I in Different Stages}

\begin{figure}[ht!]
\includegraphics[width=1\textwidth]{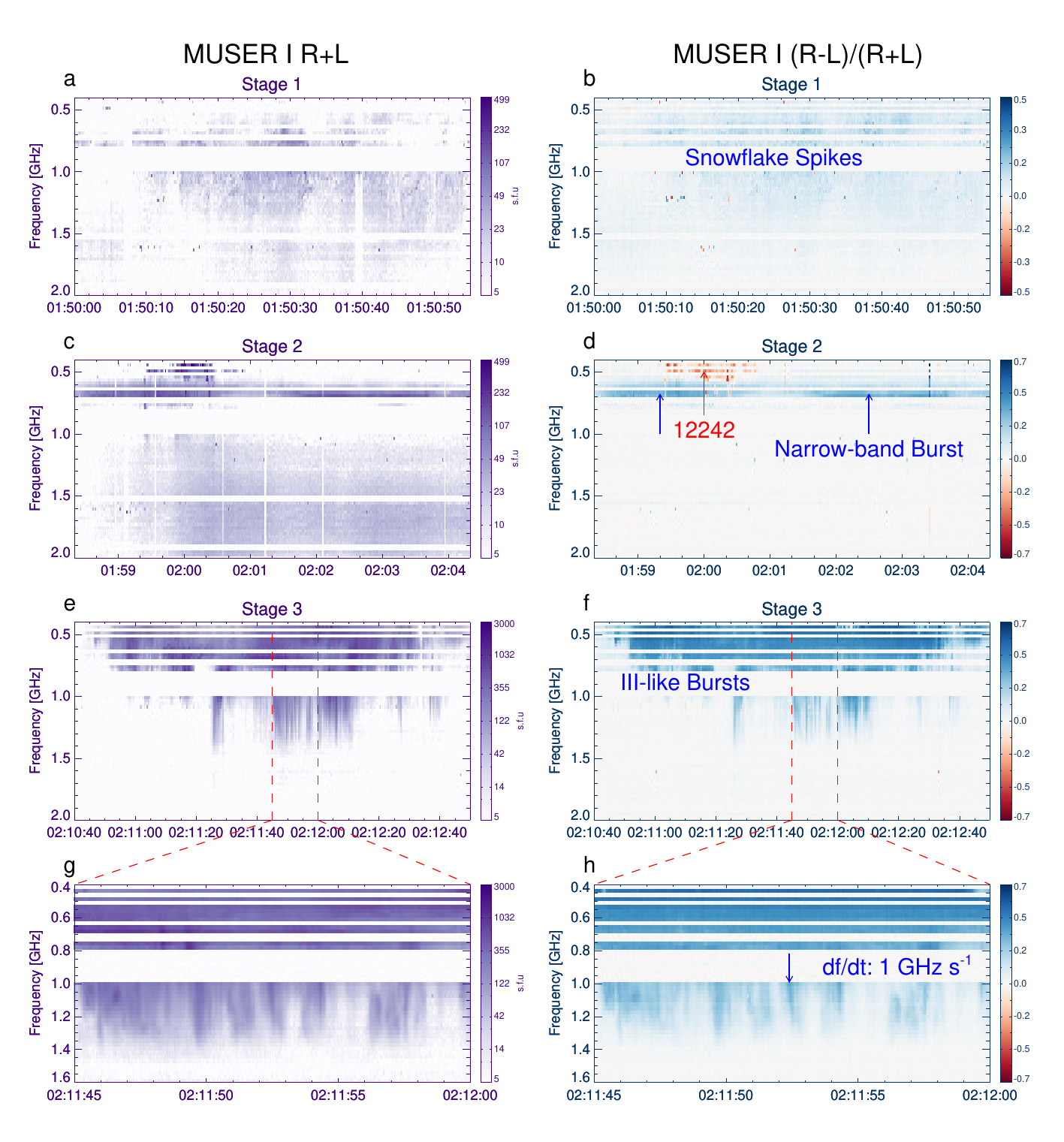}
\caption{The MUSER-I total cross-power spectrum shows fine structures in intensity (left column) and polarization (right column). Panels a/b, c/d, and e/f correspond to Stages 1, 2, and 3, respectively. Panels g and h are zoomed-in images corresponding to the time range indicated by the two red dashed vertical lines. The red arrow in panel d indicates left-polarized emission from the nearby active region AR 12242. The two blue arrows indicate narrow-band emission features. The $df/dt$ label refers to the drift rate of the structure pointed to by the blue arrow.}
\label{fig:sp_polar}
\end{figure}

Figure~\ref{fig:sp_polar} presents total cross-power spectra from MUSER-I, showing zoomed-in fine structures from Stages 1--3. The left column displays the spectral intensity given by $I_L+I_R$, while the right column shows the corresponding circular polarization, $(I_R-I_L)/(I_R+I_L)$, where $I_L$ and $I_R$ denote the intensity densities in left and right circular polarization, respectively. Panels a/b, c/d, and e/f show representative spectral characteristics of Stages~1, 2, and 3, respectively. Panels g/h provide zoomed-in views of the time interval marked by the two red dashed vertical lines in panels e/f.

During the flare peak observed at Stage~1, the emission spans a frequency range of 0.4--2~GHz, with a prominent concentration between 0.6 and 1.3~GHz. At this stage, weak right polarization is noted, with a polarization degree averaging around 0.3 and a measurement error of approximately 0.1 \citep[e.g.,][]{Yanyh_2013,Yanyh+2021}. In Stage~2, the spectrum reveals two narrow-band bursts centered at 0.67~GHz, exhibiting a bandwidth of roughly 50~MHz. These bursts are also characterized by right polarization, with the degree of polarization ranging from 0.3 to 0.4. Notably, a significant left-polarized emission detected at 02:00~UT is attributed to another active region, AR 12242.

A series of consecutive bursts occurred in Stage~3, primarily within the 0.4--0.8~GHz range, with the exception of the initial and most intense burst, which has a high-frequency cutoff at around 1.4~GHz. Panels e/f in Figure~\ref{fig:sp_polar} illustrate the most intense burst. The features observed below 0.8~GHz are likely due to the overlap of various narrow-band structures, such as narrow-band type III-like bursts and drifting structures, which may be superimposed on one another. The characteristics above 1~GHz are characterized by pulsations or bursts that resemble type III emissions. However, the gap between 0.8 and 1.0~GHz prevents further tracing of the lower section of these eruptions. A comparison of the more pronounced features at 0.8 and 1.0~GHz indicates that they do not align, suggesting that the initiation of these type III bursts likely occurs between 0.8 and 1.0~GHz. We estimated the drift rate of a reverse type III burst, which exhibits a notable drift of approximately 1~GHz~s$^{-1}$. The majority of other type III-like bursts exhibit a more notable slope, signifying a high drift rate.

The intensity of the flux density primarily ranges from 200 to 600~sfu, with a peak surpassing 1000~sfu. The degree of polarization is noted to be even greater, especially at lower frequencies, reaching around 0.5. In summary, the radio emissions captured by MUSER-I from AR 12241 between 01:40 and 02:30~UT are mainly characterized by right-polarized emissions, with a modulation degree of approximately 0.2--0.5.

\subsubsection{Source Position Evolution}

MUSER-I simultaneously observes the Sun at 64 different frequency channels. However, due to significant local interference, data from the 0.8--1.0~GHz range are unavailable, and some additional channels are unusable. Therefore, only 21 selected channels are used for imaging. For this paper, since most radio bursts linked to AR 12241 occur below 0.8~GHz and only the most intense cluster reaches a 1.4~GHz cutoff, a maximum of 10 channels is shown. Unlike the continuous and evolving signatures seen in the EUV passbands, decimetric bursts are observed as several distinct, isolated events (see Figure~\ref{fig:fig1}f).

\begin{figure}[ht!]
\includegraphics[width=1\textwidth]{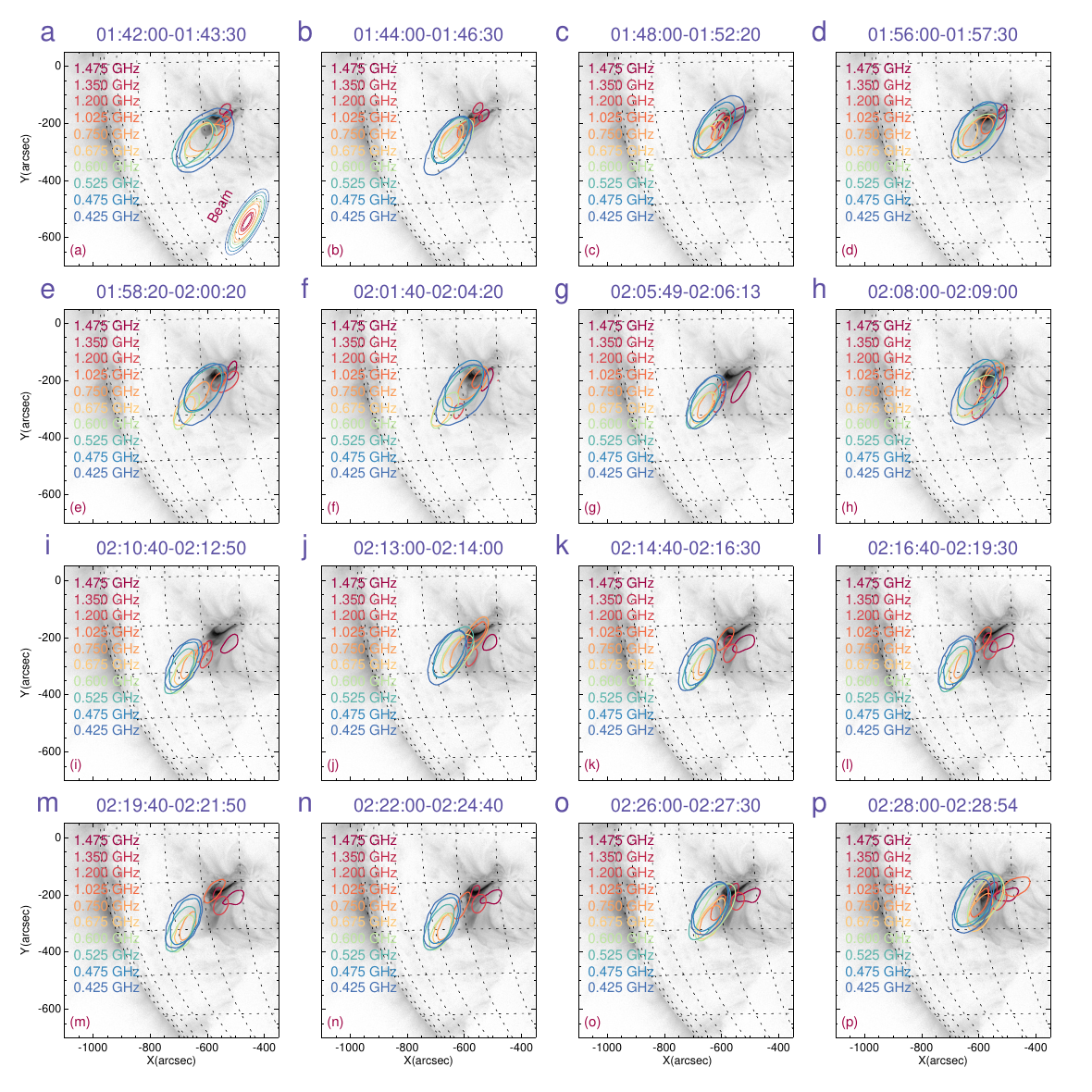}
\caption{Evolution of solar radio sources observed by MUSER-I at multiple frequencies, overlaid on AIA 94~\AA\ images. In each panel, contours indicate 90\% of the maximum intensity from the integrated radio image over the corresponding time interval, as marked in Figure~\ref{fig:fig1}f. Concentric elliptical annuli in the bottom-right corner of panel a indicate the 90\% contour of the synthesized clean beam for each frequency. The annuli colors correspond to the frequencies indicated in the legend, with each panel showing beams at its respective observation time.}
\label{fig:contour}
\end{figure}

Figure~\ref{fig:contour} presents multi-frequency radio contours overlaid on an AIA 94~\AA\ image to show the source evolution. The contours, plotted at 90\% of the individual channel's maximum intensity, use a blue-to-red color scheme to denote increasing frequency from low to high. Each panel in Figure~\ref{fig:contour} corresponds to emission integrated over a specific time interval, as denoted by the tick marks in Figure~\ref{fig:fig1}f. In the bottom right of Figure~\ref{fig:contour}a, concentric elliptical annuli show the 90\% contour of the synthesized clean beam for each frequency.

\begin{figure}[ht!]
\includegraphics[width=1\textwidth]{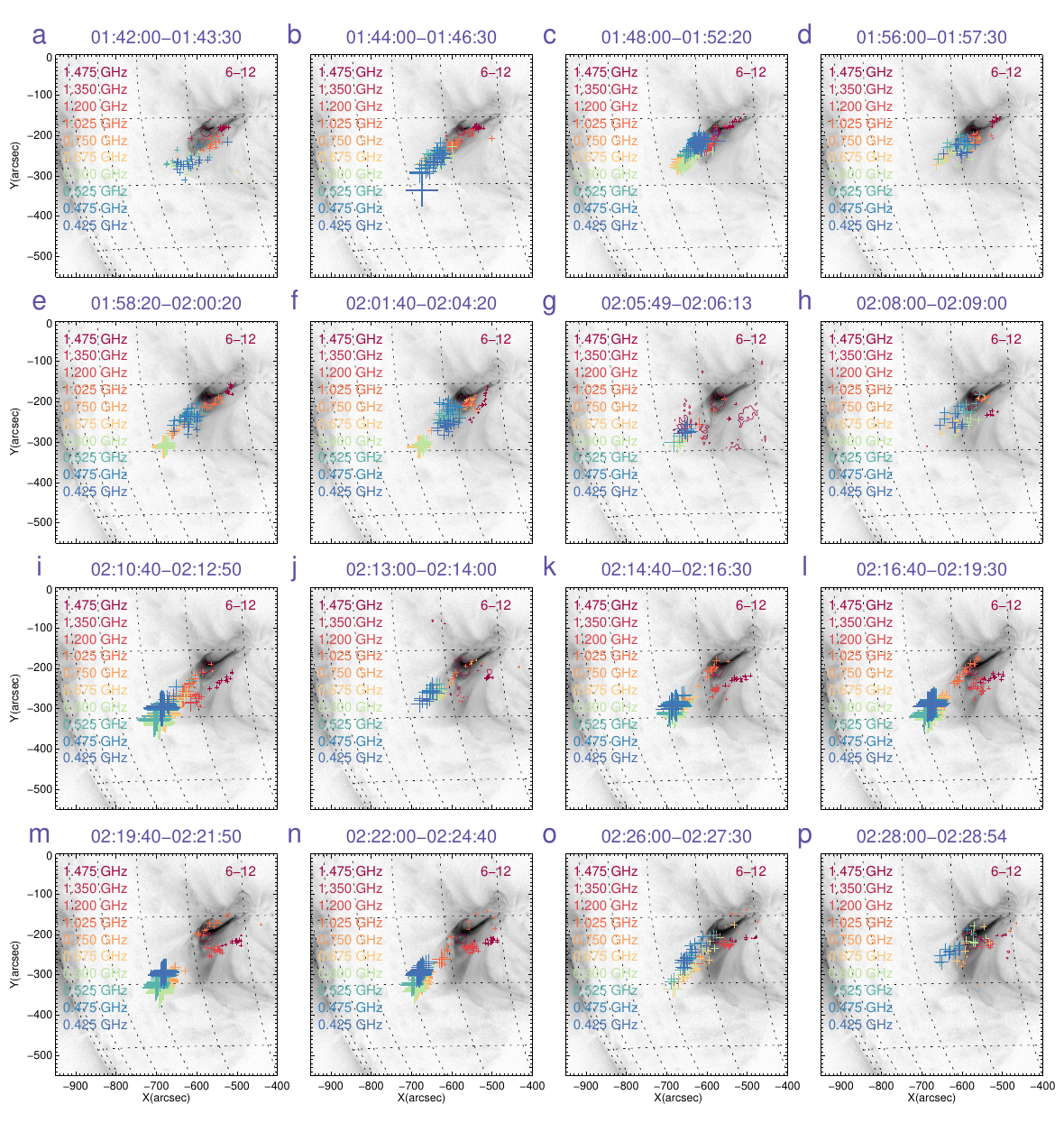}
\caption{The centroids of solar radio sources observed by MUSER-I at various frequencies, overlaid on AIA 94~\AA\ images. Panels a--p show the centroids during specific time intervals, as indicated in each panel's title and by labels `a--p' in Figure~\ref{fig:fig1}f. The size of each symbol is proportional to $2\,(\log_{10} I_{\text{peak}} - 6)$, where $I_{\text{peak}}$ denotes the peak intensity value in the corresponding image. Different colors denote different frequencies, as specified in each panel. The red contours indicate the 50\% intensity levels of RHESSI images in the 6--12~keV passband.}
\label{fig:centroid}
\end{figure}

Figure~\ref{fig:centroid} shows the centroids of the radio sources during intervals marked by horizontal lines in Figure~\ref{fig:fig1}f with a time cadence of 10~seconds, overlaid on AIA 94~\AA\ images. These centroids are determined using 2D tilted Gaussian fits to MUSER-I local images above 90\% of the peak intensity. To ensure the convergence of the Gaussian fitting, only the sources with a peak value above $1\times10^6$ (raw unit) are considered. The centroid symbols are color-coded by frequency (from blue to red). The size of each symbol is proportional to $2\,(\log_{10} I_{\text{peak}} - 6)$, where $I_{\text{peak}}$ denotes the peak intensity value in the corresponding image. The centroid uncertainty is mainly governed by phase calibration according to the method in \citet{Zhouzc+2022}, which is around one-fold the spatial angular resolution. The red contours in Figure~\ref{fig:centroid} indicate the contours of the RHESSI X-ray sources at 6--12~keV. The RHESSI emission appears mainly in the soft X-ray passband at the top of the flare arcade, changing slightly as the eruption evolves.

In Figures~\ref{fig:contour} and~\ref{fig:centroid}, panels a--c, e--g, and i, k--o show eruptions that correspond to Stages `1', `2', and `3', respectively. Phases of relative quiet are represented by panels d, h, j, and p.

As discussed in Section~3.3.1, the high-frequency component of the primary eruption is dominated by frequencies below 0.5~GHz and is represented by the prominent radio sources at lower frequencies shown in Figures~\ref{fig:contour}b and~\ref{fig:centroid}b. The radio source at the flare's peak is shown in Figures~\ref{fig:contour}c and~\ref{fig:centroid}c; these radio emissions are most likely related to electron acceleration during the inner magnetic reconnection phase. Figures~\ref{fig:sp_polar}a and~\ref{fig:sp_polar}b depict a portion of the complex spectral structure seen during this period, emphasizing the process's subtle yet dynamic nature.

Panels e--f in Figures~\ref{fig:contour} and~\ref{fig:centroid} illustrate radio sources during Stage~2. The narrow-band, more intense radio sources at 0.6--0.675~GHz are positioned at the outermost edge, distinct from those at other frequencies, maintaining a distribution similar to that observed during quiet periods. The presence of narrow-band strong emissions, along with their concentrated locations, suggests that the acceleration of electrons is likely restricted to a localized area.

Panel g in Figures~\ref{fig:contour} and~\ref{fig:centroid} depicts the sources linked to a brief eruption, as indicated by the short bar labeled `g' in Figure~\ref{fig:fig1}f. These eruptions are likely composed of multiple individual type III bursts, with sources at varying frequencies aligned along a trajectory directed toward the flare center.

Panels i, k--n in Figures~\ref{fig:contour} and~\ref{fig:centroid} display five notable bursts during Stage~3, which are likely enhanced versions of the brief eruption mentioned above, exhibiting similar spatial and spectral characteristics. The first burst is the strongest, whereas the next four eruptions are less intense, with their central frequencies probably rising as development advances, as shown in Figure~\ref{fig:fig1}f. They typically exhibit a comparable source distribution, characterized by low frequencies on the outside and high frequencies on the inside. The variation of radio sources over time will be examined in the subsequent section.

\begin{figure}[ht!]
\includegraphics[width=1\textwidth]{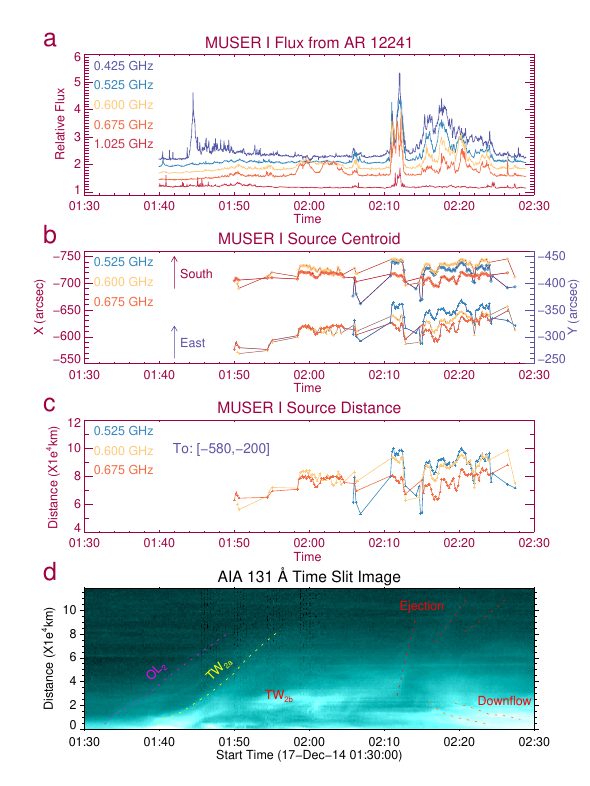}
\caption{The time profiles of MUSER-I radio flux and source centroids, together with the AIA 131~\AA\ slit image, illustrate the relationship between the dynamic evolution of the eruptive structure and the temporal variations in radio emission flux and source position.}
\label{fig:flux}
\end{figure}

\subsubsection{Temporal Relationship between Radio Emission and Eruption of \texorpdfstring{TW$_2$}{TW\_2}}

To seek the possible trigger and mechanism of the stronger radio emission, we compared the temporal relationship between radio emission and the dynamic evolution of the twisted structure. Figure~\ref{fig:flux}a presents the spatially resolved relative decimetric radio flux from AR 12241 at five frequencies. The fluxes presented here are ratio fluxes relative to the flux at 01:40~UT, which is primarily coming from the active region and is treated as background radio emission. To avoid overlap, the time profiles at different frequencies are offset vertically by increasingly larger amounts as the frequency decreases.

Figure~\ref{fig:flux}b presents the temporal evolution of the radio source centroids, while Figure~\ref{fig:flux}c shows the derived distance between the radio source centroid and the location at $[-580\arcsec, -200\arcsec]$, the location of the RHESSI source. To minimize overlap among different frequencies, only the results at 0.525, 0.600, and 0.675~GHz are shown in Figures~\ref{fig:flux}c and~\ref{fig:flux}d.

Figure~\ref{fig:flux}d presents the temporal evolution of TW$_2$ and its overlying coronal loops along the slit indicated in Figure~\ref{fig:fig2}f. At approximately 01:40~UT, the outer portion of TW$_2$, TW$_{2a}$, became apparent in the slit and exhibited a rapid rise until fading away at 01:56~UT. TW$_{2b}$ indicates the inner portion, which underwent a gradual ascent corresponding to the standard reconnection process before its final eruption at approximately 02:11~UT. During this period, narrowband bursts primarily at 0.6--0.7~GHz were detected (see Figures~\ref{fig:contour}b--c and~\ref{fig:centroid}b--c), with their source centroids situated approximately $8\times10^4$~km from the RHESSI source center. Between 02:11 and 02:14~UT, TW$_{2a}$ breaks out and appears as a bright structure in the time-slit image, highlighted by a red dashed line labeled `Ejection' in Figure~\ref{fig:flux}d.

Four subsequent clusters of radio bursts likely correspond to subsequent ejections, indicated by unlabeled red dashed lines. During this phase, both the temporal evolution of the radio source centroids and the corresponding time--distance diagrams generally follow the flux profile, exhibiting comparable trends of increase and decrease. However, after 02:20~UT, the amplitude of centroid variation deviates from this pattern: while the peak flux declines, the centroid distances continue to increase.

\section{Discussion and Conclusion}

\subsection{Main Radio Mechanism}

Based on the MUSER amplitude calibration method described in \citet{Zhouzc+2024}, the peak brightness temperature of the most intense burst was estimated to be approximately $10^7$~K, with an uncertainty of about one order of magnitude. Polarization analysis reveals that the radio bursts observed between 0.4 and 2.0~GHz were predominantly weakly to moderately right-handed, with polarization degrees typically in the range 0.2--0.5$\pm$0.1. The most prominent emission occurred at 0.5--0.8~GHz, exhibiting narrowband type III-like bursts, and extended to 1.4~GHz during the strongest burst, where the bursts showed broader bandwidth. These radio signatures are consistent with plasma emission driven by propagating electron beams. This interpretation is further supported by the particle-in-cell simulations of \citet{Yao+2022ApJ}, which show that plasma emission dominates near magnetic separatrices. Our radio sources are located along or parallel to the separatrix surface, in agreement with this scenario. The fainter radio emission is likely attributable to electron beams accelerated during reconnection that either possessed lower densities or followed a steeper energy power-law spectrum, as discussed in previous studies \citep[e.g.,][]{Libo+2014, Reid+2018b}.

\subsection{Energy Release and Particle Acceleration from Multi-wavelength Observations}

\begin{figure}[ht!]
\includegraphics[width=1\textwidth]{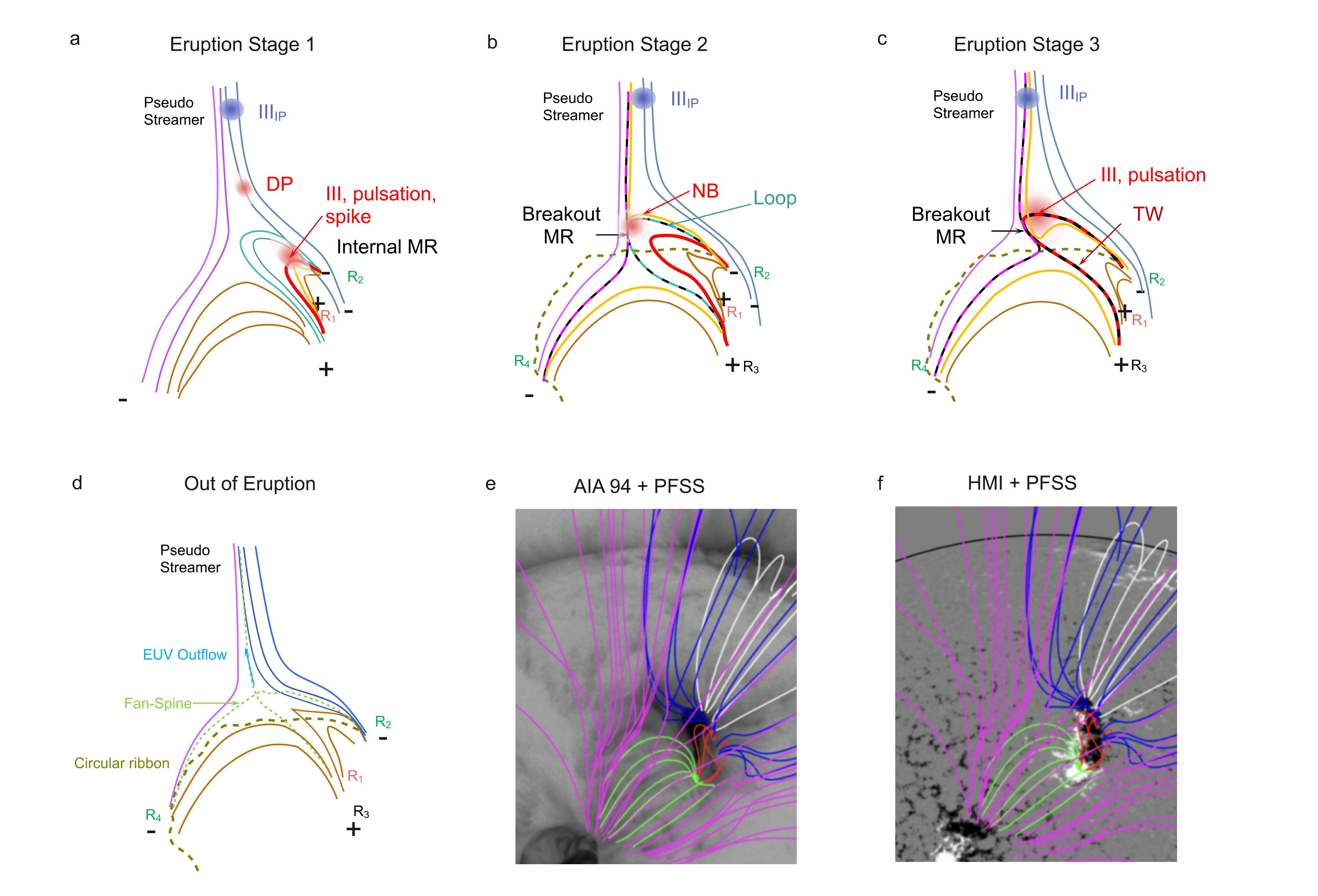}
\caption{Schematic of the generation mechanism proposed for the observed decimetric radio emission. The view is rotated clockwise by approximately 105~\textdegree to align the pseudostreamer vertically. Panels a--c depict the evolution of the magnetic topology along with the associated locations of the radio sources during eruption phases 1, 2, and 3, respectively. Panel~d presents the magnetic configuration following the eruption. Panels e and f offer a broader perspective: e features PFSS-extrapolated coronal field lines superimposed on an AIA 94~\AA\ image, while f displays the same extrapolation over an HMI line-of-sight magnetogram. The notations $R_1$--$R_4$ represent various flare ribbons, consistent with those depicted in Figure~\ref{fig:fig2}. The term ``MR'' denotes magnetic reconnection. The abbreviations ``DP'' and ``NB'' refer to drifting pulsation and narrowband burst, respectively. }
\label{fig:cartoon}
\end{figure}

By combining radio, X-ray, and EUV observations, we have constructed a coherent picture of the energy release and particle acceleration throughout the eruption. The decimetric radio bursts provide a sensitive diagnostic of nonthermal electrons accelerated by magnetic reconnection, while the soft X-ray and EUV data constrain the thermal response and the dynamic evolution of the twisted structure. Together, these multi-wavelength signatures reveal that the energy release proceeds in three distinct stages, each characterized by different acceleration sites and particle propagation patterns, as illustrated in Figure~\ref{fig:cartoon}.

Recent studies have demonstrated that magnetic null points and quasi-separatrix layers are key sites for electron acceleration in solar eruptions \citep[e.g.,][]{Dudik+2025}. In particular, \citet{Kumar+2025} showed that in nested-null topologies beneath a pseudostreamer, two successive breakout reconnections at inner and outer null points are responsible for the onset of CMEs and the quasiperiodic injection of electron beams. Our observations are consistent with this framework: the decimetric bursts are spatially associated with high-\(Q\) regions derived from NLFFF extrapolations, indicating that acceleration occurs along the QSLs.

Panels a--c of Figure~\ref{fig:cartoon} depict the evolution of the magnetic topology and the corresponding radio source locations during Stages~1--3 as defined in Section~3.1. Panel~d illustrates the relaxed post-eruption morphology, which exhibits a pseudostreamer and an incomplete fan-spine configuration. Panels e and f display PFSS-extrapolated magnetic field lines superimposed on an AIA 94~\AA\ image and an HMI line-of-sight magnetogram, respectively.

The purple and blue lines trace the outer magnetic field emanating from a negative-polarity photospheric area, which, together with the underlying coronal arcades, forms the pseudostreamer. The thick red line denotes the twisted structure. Turquoise solid lines indicate the overlying coronal loops. The turquoise-black and purple-black dashed curves mark the closed and open magnetic loops undergoing magnetic reconnection. After reconnection, newly formed loops and curved field lines appear in gold. The notations $R_1$--$R_4$ represent various flare ribbons, consistent with those depicted in Figure~\ref{fig:fig2}.

In Stage~1 (Figure~\ref{fig:cartoon}a), the energy release is dominated by internal magnetic reconnection associated with the internal sheared twisted loops, corresponding to the main phase of the flare. Below 500~MHz, a series of drifting pulsations (``DP'') and interplanetary type III bursts (``III$_\mathrm{IP}$'') are observed, indicating that the accelerated electrons propagate upward. These drifting pulsation structures are widely interpreted as signatures of plasmoids formed during magnetic reconnection \citep[e.g.,][]{Karlicky+2025}, while the interplanetary type III bursts reflect electron beams escaping along open field lines into the heliosphere \citep{Paipaleon+2025}. Above 500~MHz, spikes, pulsations, or type III bursts appear, indicating that the accelerated electrons are produced near the cusp region or propagate downward. This scenario is consistent with the standard flare model, in which electrons are accelerated in the reconnection region and produce the observed radio bursts. The coexistence of pulsations and spikes suggests that the acceleration is intermittent and dynamic, likely modulated by the fragmented nature of the internal reconnection \citep[see][]{Clarke+2021, LiD+2024}.

In Stage~2, two narrow-band radio bursts, each lasting approximately two minutes, were observed, confined primarily to the 0.6--0.8~GHz band. The radio source was located near the site labeled ``NB'' in Figure~\ref{fig:cartoon}b, where the QSL was located (labeled ``Breakout MR''). These results suggest that electrons are locally accelerated by magnetic reconnection at QSL between the overlying loops (turquoise-black dashed curves) and the open field lines rooted in the negative magnetic field (purple-black dashed curves). This behavior is consistent with recent reports of quasiperiodic decimetric type III bursts arising during inner breakout reconnection in nested-null topologies \citep{Kumar+2025}.

In Stage~3 (Figure~\ref{fig:cartoon}c), the most intense decimetric bursts occur, corresponding to the eruption of the main body of TW$_2$; however, no clear thermal enhancement is identified. Most radio bursts have a high-frequency cutoff below 1.4~GHz. In the context of plasma emission, this indicates that electron acceleration occurred above a certain altitude and did not extend into the low corona. Based on the EUV observations and the derived \(\log Q\) distribution, the nonthermal electrons are likely produced by breakout magnetic reconnection between TW$_2$ (indicated by a thick red-black dashed curve) and adjacent open field lines (represented by a purple-black dashed line). Betatron/slingshot acceleration from the restoring magnetic tension of the newly formed, highly curved field lines likely provides an additional acceleration mechanism or guides the propagation of electrons.

\subsection{Why Stronger Decimetric Emissions Occur in the Decay Phase}

Our analysis reveals a marked contrast in the radio emission signatures detected by MUSER-I (0.4--2~GHz) across different stages. The decimetric emission during Stage~1, the impulsive and peak phase of the flare, is notably weak, whereas the emission during Stage~3, the deep decay phase, is significantly more intense. This difference can be understood primarily in terms of the acceleration geometry associated with the two reconnection sites.

In Stage~1, the energy release is dominated by internal magnetic reconnection rooted at relatively low altitudes. The diffusion region is compact, the reconnection electric field is spatially limited, and the geometry does not favor the production of field-aligned electron beams. Accelerated electrons that travel downward encounter dense chromospheric plasma and lose their energy through collisions and free--free absorption before producing significant radio emission. Electrons that travel upward are few and are quickly scattered. As a result, the decimetric emission is weak, and no hard X-ray signature is detected.

In Stage~3, the situation is fundamentally different. The breakout reconnection occurs at higher altitude, at an extended high-squashing-factor \(Q\) region or separatrix surface between the outer envelope of the erupting structure and the surrounding open field lines. This geometry is favorable for electron acceleration in several ways. First, the diffusion region is larger, so a greater number of electrons can be accelerated per unit time. Second, the reconnection electric field is stronger and more extended along the separatrix surface, which efficiently accelerates electrons through curvature drift and parallel electric fields \citep{Kumar+2025,Xia+2020}. Third, the accelerated electrons are preferentially directed upward along open field lines into the tenuous outer corona, where both free--free and gyroresonance absorption are weak. This allows the coherent plasma emission to escape efficiently, producing strong decimetric bursts.

The repeated sub-eruptions in Stage~3 further enhance the emission. Each sub-eruption injects a fresh beam of nonthermal electrons into the same high-\(Q\) region. The superposition of these beams along the narrow separatrix layer amplifies the plasma emission, leading to the observed successive clusters of type III-like bursts. This is consistent with the quasiperiodic nature of breakout reconnection in nested-null topologies \citep{Kumar+2025, Karlicky+2025}.

Stage~3 lacks a clear hard X-ray signature, consistent with most accelerated electrons escaping upward along open field lines or losing energy before reaching the chromosphere, thereby producing no detectable thick-target bremsstrahlung. In contrast, Stage~1 internal reconnection is rooted closer to the chromosphere, allowing downward precipitation despite the modest energy release. These factors together account for the strong successive decimetric bursts observed during the decay phase.

\subsection{Conclusion}

This study presents a comprehensive analysis of decimetric radio bursts associated with a complex solar eruption beneath a coronal pseudostreamer, using high-cadence radio imaging from MUSER-I and multiwavelength data from SDO, NoRH, and RHESSI. Our results demonstrate the following.

First, the radio emission during the impulsive phase of the M1.1 flare is weak, compared with the decay phase. The characteristics of the bursts during the impulsive phase, including drifting pulsations and narrowband spikes, are consistent with the standard flare model in which electrons are accelerated in the reconnection region underlying the erupting magnetic structure. The coexistence of pulsations and spikes indicates that the acceleration is intermittent and dynamic, likely modulated by the fragmented nature of internal reconnection \citep{Clarke+2021, LiD+2024}. The disappearance of some internal twisted loops during this period further supports the interpretation that internal reconnection governs the energy release at this stage.

Second, the strongest decimetric emission occurs during the deep decay phase of the flare, manifesting as several clusters of pulsations. This emission is likely caused by electron acceleration during breakout magnetic reconnection associated with the main body of the twisted structure TW$_2$. The radio source locations coincide with high-\(Q\) regions derived from NLFFF extrapolations, which are indicative of QSLs and preferred sites for magnetic reconnection \citep{Dudik+2025}. This spatial association provides evidence consistent with electron acceleration being driven by breakout reconnection rather than by the standard flare current sheet.

The observed decimetric bursts provide compelling evidence for electron acceleration via separator reconnection in a pseudostreamer environment. Recent studies of nested-null topologies have shown that quasiperiodic pulsations in radio and EUV emissions are commonly observed during the interaction of erupting flux ropes with overlying magnetic fields through breakout reconnection near both inner and outer nulls \citep{Kumar+2025}. The reconnection is inherently bursty and quasiperiodic, modulated by plasmoids or fast MHD waves. The existence of two breakout current sheets in such nested-null configurations provides additional opportunities for particle acceleration, remote footpoint heating, and particle escape into the heliosphere through reconnection at the outer current sheet \citep{Kumar+2025}. Our results are consistent with this framework and further demonstrate that the concentrated source positions of the strong successive radio bursts suggest that pseudostreamers may act as effective particle accelerators, trapping electrons and directing them along a narrow, well-defined layer. These findings highlight the unique diagnostic value of decimetric radio imaging for tracing particle acceleration and magnetic connectivity in the dynamic solar corona.

\begin{acknowledgments}
This work is supported by the National Key R\&D Program of China (NKRDPC) under Grants No.~2021YFA1600503 (2021YFA1600500) and No.~2022YFF0503800, and by the Specialized Research Fund for the State Key Laboratory of Solar Activity and Space Weather (SKLSASW). X.Z. was also supported by NKRDPC No.~2025YFE0202500, the open research program SASW-2025KFKT06-3 of the SKLSASW, and China's Space Origins Exploration Program (GJ11020405). H.R. acknowledges the UKSA grant UKRI980, ST/X002012/1, and the STFC grant ST/W001004/1. H.C. was supported by NKRDPC No.~2021YFA1600502 (2021YFA1600500) and the Strategic Priority Research Program of the Chinese Academy of Sciences, Grant No.~XDB 41000000. X.C. was supported by NKRDPC No.~2025YFF0510700.

We acknowledge the Chinese Meridian Project (https://cstr.cn/31122.02.MSP) for supporting the operation of MUSER. We also thank the MUSER team, the SDO/AIA team, and the SDO/HMI team for the data.
\end{acknowledgments}

\begin{contribution}
S.M. led the project, performed the data analysis, and wrote the manuscript.
Y.Y. supervised the project, acquired funding, and contributed to interpretation.
W.W. contributed to the foundational MUSER data-processing code and provided guidance on calibration.
H.R. contributed to the interpretation of radio emission and edited the manuscript.
T.L. contributed to data analysis and interpretation, and edited the manuscript.
X.Z. performed the NLFFF extrapolations and QSL analysis.
H.C. performed the NLFFF extrapolations and data analysis, and edited the manuscript.
Z.Z. contributed to MUSER imaging calibration.
J.H. contributed to radio data analysis and interpretation.
L.C. contributed to MUSER image calibration.
C.T. contributed to MUSER spectral calibration.
X.C. contributed to the discussion of radio emission.
M.Z. contributed to the discussion of radio emission.
X.Y. contributed to the discussion of electron acceleration.
All authors discussed the results and commented on the manuscript.
\end{contribution}
\bibliography{ms_12241_main}
\bibliographystyle{aasjournalv7}

\end{CJK*}

\end{document}